\documentclass[final,5p,times,twocolumn
]{elsarticle}

\usepackage{graphicx}
\usepackage{dcolumn}
\usepackage{bm}
\usepackage{amsmath,amssymb,amsfonts}
\usepackage{hyperref}

\DeclareMathAlphabet{\pazocal}{OMS}{zplm}{m}{n}

\def\imo{i}

\def\K{{\cal K}}

\begin{document}
\begin{frontmatter}

\title{Black holes in Starobinsky-Bel-Robinson Gravity and the breakdown of quasinormal modes/null geodesics correspondence}

\author[first]{S. V.  Bolokhov}
\affiliation[first]{organization={Peoples' Friendship University of Russia (RUDN University)},
            addressline={6 Miklukho-Maklaya Street}, 
            city={Moscow},
            postcode={117198}, 
            country={Russia}}

\begin{abstract}
We show that perturbations of a scalar field in the background of the black hole obtained with the  Starobinsky-Bel-Robinson Gravity is unstable unless   the dimensionless coupling $\beta$ describing the compactification of M-theory is small enough. In the sector of stability quasinormal spectrum show peculiar behavior both in the frequency and time domains: the ringing consists of two stages where two different modes dominate. The WKB method does not reproduce part of the spectrum including the fundamental mode, which is responsible for the first stage of the ringing. As a result, the correspondence between the high frequency quasinormal modes and characteristics of the null geodesics  reproduces only one branch of the eikonal spectrum. The frequencies are obtained with the help of three methods (Frobenius, WKB and time-domain integration) with excellent agreement among them. 
\end{abstract}


\end{frontmatter}

\section{\label{sec:Introduction}Introduction}

Quasinormal modes (QNMs) of black holes are their fundamental characteristics, proper frequencies of oscillations which depend on the parameters of the black hole and not on the way in which a black was perturbed \citep{Kokkotas:1999bd,Nollert:1999ji,Konoplya:2011qq,Berti:2009kk}.  
The data from gravitational waves together with observations in the electromagnetic spectra \citep{Barack:2018yly,EventHorizonTelescope:2022xqj,Goddi:2016qax} in principle should allow one in the future to determine the black hole's mass, angular momentum, and charge and thereby to test gravitational theory in the strong field limit. 
Nevertheless, by now the uncertainty in the mass and angular momentum of a black hole is large, what allows for big room for alternative theories of gravity \citep{Konoplya:2016pmh,Yunes:2016jcc}.

Among various alternative/modified theories of gravity the Starobinsky inflation model plays an important role. It uses the idea that the early universe went through an inflationary de Sitter era \citep{Starobinsky:1979ty} and was conceived by Alexei A. Starobinsky in \citep{Starobinsky:1980te}. The model incorporates modifications to General Relativity by introducing a quadratic curvature term. This framework allows for an inflationary phase driven by a scalar field, addressing cosmological issues like the horizon and flatness problems. It generates a scale-invariant density perturbation spectrum, consistent with cosmic microwave background and large-scale structure observations, making it a significant inflationary model in modern cosmology. 

Recently a new gravitational theory in four dimensions has been proposed in the form of
a sum of the modified $R + \alpha R^2$  Starobinsky inflation allowing for the leading
Bel–Robinson-tensor-squared term. These corrections are inspired by the gravitational effective action of superstrings/M-theory
compactified to four dimensions  \citep{Ivanov:2021chn,Ketov:2022lhx,CamposDelgado:2022sgc,Ketov:2022zhp}.  
The Starobinsky–Bel–Robinson Lagrangian possesses only two free parameters, which
makes it attractive for studying various physical phenomena which could potentially be observed and, thereby, the theory could be tested. Consequently, the corrected Schwarzschild-like spacetime was perturbatively obtained in \citep{CamposDelgado:2022sgc}, while cosmological applications were considered in \citep{Ketov:2022zhp,Do:2023yvg,Pozdeeva:2023djw}. Various effects around such black holes have been recently considered in \citep{Belhaj:2023dsn,Arora:2023ijd}

Here we propose the first study of the quasinormal frequencies in the background of  black holes in the Starobinsky–Bel–Robinson gravity. 
We will consider a test scalar field, which is the simplest case  frequently sharing the properties of other spin fields. 
Apparently the most important aspect of the quasinormal spectrum is its (in)stability \citep{Whiting:1988vc,Dias:2020ncd,Ishihara:2008re,Kodama:2009bf,Cardoso:2009cnd,Takahashi:2010gz,Cuyubamba:2016cug,Zhu:2014sya}, because usually it is difficult  to prove the stability analytically, and the thorough analysis of quasinormal modes allows one to judge about (in)stability. The onset of instability in the scalar sector may indicate the  possible instability in the gravitational one. The neutral test scalar field in the Schwarzschild, Kerr and Kerr-Newman spacetimes are known to be stable under the quasinormal mode boundary conditions, because the corresponding quasinormal frequencies are damped \citep{Ohashi:2004wr,Konoplya:2006br,Konoplya:2013rxa}.

Here we will consider the evolution of perturbations of the scalar field in the black hole background in the Starobinsky–Bel–Robinson gravity.
We will show that the scalar field is unstable unless the higher curvature correction parameter $\beta$ is small enough. We find the critical value of the parameter $\beta$ at the threshold of instability and analyze the quasinormal spectrum in the stable sector. The latter has distinctive features: the ringdown consists of the two stages at which two different modes dominate. One of them cannot be reproduced by the WKB method what undermines the correspondence between eikonal quasinormal modes and null goedesics.

The paper is organized as follows. In sec. II we briefly describe the black hole spacetime and wave-like equation. Sec. III summarizes the methods used for the analysis of the quasinormal spectrum, while in sec. IV we present the obtained results for the instability threshold and quasinormal modes in the range of parameters corresponding to the stability. In the Conclusions we review the obtained results and mention some points for future investigation. 

\section{The metric}

Using the compactification coupled with the presence of stringy fluxes, the appropriate $4D$ gravity models may be described by the following action:
\begin{equation}
S=\frac{M_{p l}}{2} \int d^{4} x \sqrt{-g}\left(R+\frac{R^{2}}{6 m^{2}}-\frac{\beta}{32 m^{6}}\left(P_{4}^{2}-E_{4}^{2}\right)\right).
\label{1}
\end{equation}
Here $g$ is the metric determinant, $R$ is the Ricci scalar; $m$ is a free mass parameter, $\beta>0$ is a dimensionless coupling describing the compactification of M-theory, $P_{4}^{2}$ and $E_{4}^{2}$ are the Pontryagin and Euler topological densities which are related to the Bel-Robinson tensor $T_{\mu \nu \lambda \rho}$ in four dimensions as follows:
\begin{equation}
T^{\mu \nu \delta \rho} T_{\mu \nu \delta \rho}=\frac{1}{4}\left(P_{4}^{2}-E_{4}^{2}\right).
\label{2}
\end{equation}
The Bel-Robinson tensor has the following form:
\begin{equation}
T^{\mu \nu \delta \sigma}=R^{\mu \rho \gamma \delta} R_{\rho \gamma}^{\nu \sigma}+R^{\mu \rho \gamma \sigma} R_{\rho \gamma}^{\nu \delta}-\frac{1}{2} g^{\mu \nu} R^{\rho \gamma \alpha \delta} R_{\rho \gamma \alpha}^{\sigma}.
\label{3}
\end{equation}
It can be seen from Eq.(\ref{1}) that the gravity action depends only on the two parameters, $m$ and $\beta$. This allows for various applications such as Hawking radiation, entropy, inflation, optical phenomena etc. \citep{Ivanov:2021chn,Ketov:2022lhx,CamposDelgado:2022sgc,Ketov:2022zhp,Arora:2023ijd} The black hole solution in SBR gravity depends on the value of $\beta$ corrected to first order perturbations.

The line element of the spherically symmetric metric is
\begin{equation}
d s^{2}=-f(r) d t^{2}+\frac{1}{f(r)} d r^{2}+r^{2} d \Omega^{2}.
\label{4}
\end{equation}
where the metric function has the form \citep{CamposDelgado:2022sgc}:
\begin{equation}
f(r)=1-\frac{2M}{r}+\beta M^6 \left(\frac{8 \sqrt{2} \pi  M }{r^{3}}\right)^{3}\left(\frac{108 r-194 M}{5 r}\right).
\label{5}
\end{equation}
Here $r_{s}=2 M$ is the Schwarzschild radius and $M$ is the mass parameter. The tidal force effects and shadows for this metric were considered in \citep{Arora:2023ijd}.

Coefficient $(8 \sqrt{2} \pi)^3 \approx 44901.9$ which means that $\beta \sim 3 \cdot 10^{-5}$ (at $M=1$) must already be considered as large deformation of the Schwarzschild spacetime.
As we are interested here also in the testing of the null geodesics/eikonal quasinormal modes correspondence correspondence for generic black-hole spacetimes, we will not be limited by the regime of tiny $\beta$ only and include consideration of relatively large deformations of the Schwarzschild spacetime which already cannot be described as a small correction.

\begin{figure*}
 \resizebox{\linewidth}{!}{\includegraphics{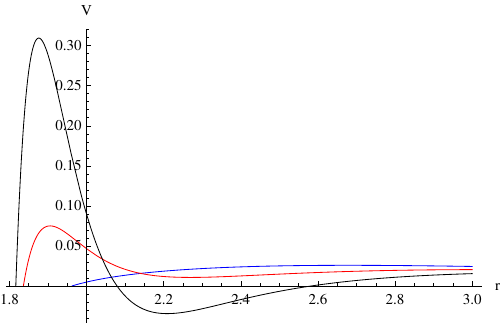}\includegraphics{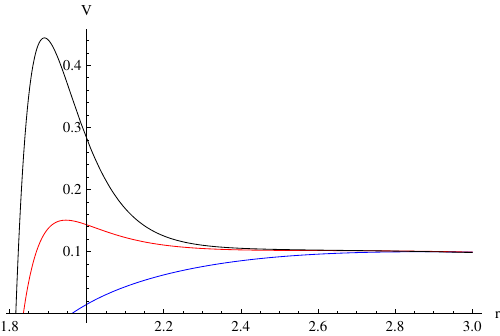}}
\caption{Effective potentials for $\ell=0$ (left) and $\ell=1$ (right) at $\beta =10^{-4}$ (blue),  $\beta =10^{-3}$ (red), $\beta = 0.002$ (black); $M=1$.}\label{figPot}   
\end{figure*}

\section{The methods}

\subsection{Wave-like equation and boundary conditions}

We will consider the test neutral massless scalar field $\Phi$ which obeys the general covariant Klein-Gordon equation,
\begin{equation}\label{K-G}
    \frac{1}{\sqrt{-g}}\partial _{\mu }\left(g^{\mu\nu}\sqrt{-g}\partial_{\nu}\Phi \right)=0.
\end{equation}
The separation of variables could be performed using the stationary ansatz,
\begin{equation}\label{separation1}
    \Phi=\frac{\psi_{\ell}(r)}{r}Y_{\ell m}(\theta ,\phi ) e^{-i\omega t},
\end{equation}
which transforms the Klein-Gordon equation to the wave-like form:
\begin{equation}\label{radial2}
    \frac{d^2 \psi_{\ell}}{d r_{*}^2} +\left [  \omega ^{2}-U(x) \right ]\psi_{\ell}=0.
\end{equation}
The effective potential is
\begin{equation}\label{potential}
    U(r)= f(r) \left(\frac{\ell(\ell+1)}{r^2} + \frac{ f'(r)}{r}\right),
\end{equation}
where the tortoise coordinate $x$ can be expressed as follows:
\begin{equation}\label{toordinate}
    \dfrac{d r_{*}}{d r}=\frac{1}{f(r)}.
\end{equation}
The effective potential is shown on fig \ref{figPot}. There one can see that the when $\beta$ is increasing, the additional gap appears, which at sufficiently large values of $\beta$
becomes negative. This is an indication that a bound state with negative energy may be formed, which means the onset of instability. Therefore, thorough analysis of quasinormal modes is necessary. 

The quasinormal modes are eigenvalues of the above second order differential equation under the requirements of purely ingoing waves at the event horizon and purely outgoing waves at infinity. 

\subsection{Higher order WKB method}

For an additional check of accurate results obtained by the Frobenius method  in the range of stability we will also use the WKB method. It was applied for the first  time to the problem of finding of quasinormal modes and grey-body factors of black holes by B. Schutz and C. Will \citep{Schutz:1985km}. The first order WKB formula was afterwords generalized to higher orders up to the 13-th order  \citep{Iyer:1986np,Konoplya:2003ii,Matyjasek:2017psv}. Recently it has been further improved by using the Pad\'e approximants \citep{Matyjasek:2017psv}, so that the relative error of the WKB formula with Pad\'e approximants is usually from quite  few times to a few orders smaller than without them.
The WKB formula can be expressed in the following general form \citep{Konoplya:2019hlu}:
\begin{eqnarray}\label{WKBformula-spherical}
\omega^2&=&U_0+A_2(\K^2)+A_4(\K^2)+A_6(\K^2)+\ldots\\\nonumber&-&\imo \K\sqrt{-2U_2}\left(1+A_3(\K^2)+A_5(\K^2)+A_7(\K^2)\ldots\right),
\end{eqnarray}
where for the quasinormal frequencies we have
\begin{equation}
\K=n+\frac{1}{2},  \quad n=0,1,2,\ldots
\end{equation}
Here $n$ is the overtone number, $U_0$ is the value of the potential's maximum, $U_2$ is the value of the second derivative of the potential in this point, and $A_2,A_3,A_4,\ldots$ depend on higher derivatives of the potential in its maximum.  In the present work we apply the sixth order WKB method of \cite{Konoplya:2003ii}, but with Pad\'e approximants  $\tilde{m}=4$,  as prescribed in \citep{Matyjasek:2017psv,Konoplya:2019hlu}. This kind of WKB approach is frequently and effectively used for calculations of quasinormal frequencies and greybody factors and usually shows good agreement with the precise methods (see, for example \citep{Matyjasek:2021xfg,Guo:2023ivz,Matyjasek:2020bzc,Konoplya:2019xmn}).

\subsection{The method of integration in time-domain}

The integration in the time domain that we employ as the main method here is founded on the wavelike equation expressed in terms of the light-cone variables $u=t-r_*$ and $v=t+r_*$. We implement the discretization approach introduced by Gundlach, Price and Pullin \citep{Gundlach:1993tp}:
\begin{eqnarray}
\Psi\left(N\right)&=&\Psi\left(W\right)+\Psi\left(E\right)-\Psi\left(S\right) \nonumber \\ \label{Discretization}
&-&\Delta^2U\left(S\right)\frac{\Psi\left(W\right) + \Psi\left(E\right)}{4}+{\cal O}\left(\Delta^4\right)\,
\end{eqnarray}

Here, the notations for the points are the following:
$N\equiv\left(u+\Delta,v+\Delta\right)$, $W\equiv\left(u+\Delta,v\right)$, $E\equiv\left(u,v+\Delta\right)$, and $S\equiv\left(u,v\right)$. 
We assume that an initial Gaussian wave package on the surfaces $u=u_0$ and $v=v_0$ model the perturbation. The obtained quasinormal frequencies are known to be independent on the parameters of this package, such as its height and location of the the center. The quasinormal frequencies can then be found from the time-domain profiles with not very high but guaranteed accuracy using the representation of the signal as a sum of exponents which is the Prony method, as detailed, for instance, in ~\citep{Konoplya:2011qq}.  As a rule one can find quasinormal modes with the relative error of less than one percent at the first or higher multipoles \citep{Bronnikov:2019sbx,Konoplya:2020jgt}, but not always at the $\ell=0$ case, because the time of the ringing is very short for that case.
The challenging point of the method is the construction of the effective potential as a function of the tortoise coordinate $r^{*}$ which can be done in the following way. One can numerically integrate the equation $d r^{*} =dr/f(r)$ and find $r^{*}(r)$ then using its connection with the coordinates $u,v$ to built $U(r*)$. An important feature of the time-domain integration which we will use here is that it includes contributions of all overtones and, is, thereby, efficient for detecting the instability. 

\begin{figure*}
 \resizebox{\linewidth}{!}{\includegraphics{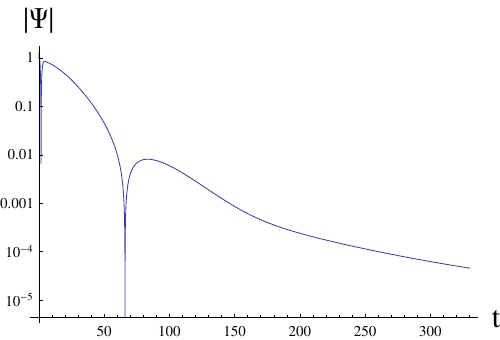}\includegraphics{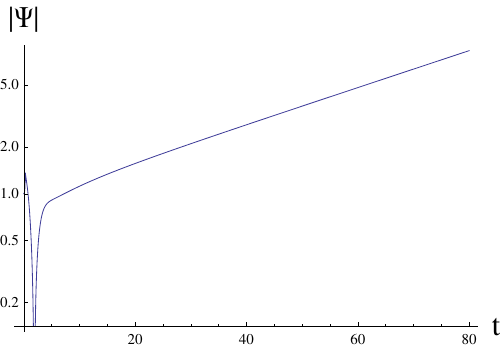}}
\caption{Time domain profiles for the stable decayed oscillations $\beta = 0.13635$ (left) and at the  instability $\beta = 0.13640$ (right); $M=1$, $\ell=0$.}\label{figInstability}   
\end{figure*}

\subsection{Frobenius method}

The Frobenius or Leaver method is based on the convergent procedure and is therefore allows us to find quasinormal frequencies with great accuracy ~\citep{Leaver:1985ax,Leaver:1986gd}. As it is also broadly discussed in the literature, here we will only briefly mention the key points of the method. The differential master wave-like equation has a regular singular point at the event horizon $r=r_g$ and the irregular singular point at $r=\infty$. Using the new function of the radial coordinate $P(r)$,
\begin{equation}\label{reg}
\Psi(r)= P (r, \omega) \left(1-\frac{r_g}{r}\right)^{-\imo\omega/F'(r_g)}y(r),
\end{equation}
we then choose the factor $P$  such  that it provides regularity if $y(r)$ in the range $r_0\leq r<\infty$, for the outgoing wave at  infinity and ingoing wave at the horizon. Therefore, we can write down the Frobenius series in the following way:
\begin{equation}\label{Frobenius}
y(r)=\sum_{k=0}^{\infty}a_k\left(1-\frac{r_g}{r}\right)^k.
\end{equation}
Then, the Guassian eliminations reduce the problem to an algebraic equation. In addition,  for further increasing of the convergence we will use the Nollert improvement \citep{Nollert:1993zz}, which was generalized to an arbitrary number of terms recurrence relations by A. Zhidenko in \citep{Zhidenko:2006rs}. The Frobenius method, however, requires an initial guess for the value of quasinormal frequency, when finding the roots of the corresponding algebraic equations and if the corresponding zeros are highly localized in the parametric space, it might be difficult to find such an initial guess for every mode, so that there is a risk of missing some modes. Therefore the WKB and time-domain integration methods are useful for getting such an initial guess.  

\section{The threshold of instability}

From the form of the effective potentials it is evident that once $\beta$ is increased, the negative gap outside the black hole becomes deeper (fig. \ref{figPot}), so that one should check whether the scalar field is stable. Time-domain integration at various values of $\beta$ and $\ell=0$ shows that, indeed, the instability occurs at some critical value:
\begin{equation}
\beta_{cr} \approx 0.1364.
\end{equation}

Although usually perturbations at higher multipole numbers are more stable, because the effective potential get higher, there are examples of instability developing at high $\ell$ as well \citep{Takahashi:2010gz,Konoplya:2017lhs}, so that higher multipoles must be investigated for possible (in)stability as well. The modes with $\ell=1, 2$ and higher multipoles (see, for example, table II) are stable at least for those values of $\beta$ for which the stability of the $\ell=0$ perturbations is guaranteed. The unstable mode has non-oscillatory exponential form as can be seen from fig. \ref{figInstability}. This non-oscillatory character of the unstable mode was rigorously proved in \citep{Konoplya:2008yy} for spherically symmetric black holes.  

Because of the large numerical pre-factor, the threshold of instability corresponds to a very large deformation of the Schwarzschild spacetime, so that, obviously, the regime of small $\beta$-corrections is safe as to the possible development of instability.

\section{Quasinormal frequencies in the stable sector}

\begin{table}
\begin{tabular}{ccc}
\hline
 $\beta$ & \text{Time-domain}  & \text{WKB} \\
\hline
$10^{-6}$ &           0.110022 - 0.105870 i  &  0.109460 - 0.103417  i  \\
$5 \cdot 10^{-6}$ &   0.110119-  0.105857 i    &  0.109426 - 0.103701  i  \\
$10^{-5}$ &   0.110244 - 0.105843 i   &   0.109436 - 0.103135 i  \\
$3 \cdot  10^{-5}$ &  0.110766 - 0.105798 i   &   0.083291 - 0.094837  i  \\
$5 \cdot  10^{-5}$ &  0.111322 - 0.105778 i   &   0.091824 - 0.129149 i  \\
$7 \cdot  10^{-5}$ &  0.111892 - 0.105784 i   &   0.086486 - 0.138424 i    \\
$10^{-4}$ &           0.112733 - 0.105896 i  &   0.032663 - 0.143389 i  \\
$10^{-3}$ &          0.121337 - 0.115261 i &     -- \\
\hline
\end{tabular}
\caption{Quasinormal modes obtained by the time-domain integration and subsequent extraction of frequencies with the Prony method for various values of $\beta$, $\ell=0$, $M=1$. The higher order WKB data with Pad\'e approximants are evidently diverge from the correct time-domain integration results when $\beta$ grows.}
\end{table}

\begin{table}
\begin{tabular}{ccc}
\hline
 $\beta$ & \text{Time-domain}  & \text{WKB} \\
\hline
$10^{-6}$ &           0.293003 - 0.097665 i  &   0.292990 - 0.097690 i  \\
$5 \cdot 10^{-6}$ &   0.293090 - 0.097568 i    &   0.293246 - 0.097705 i  \\
$10^{-5}$ &   0.293202-  0.097447 i   &   0.293569 - 0.097658 i  \\
$2 \cdot  10^{-5}$ &  0.293430-  0.097203 i   &   0.294184 - 0.097428 i  \\
$3 \cdot  10^{-5}$ &  0.293667-  0.096956 i   &   0.294743 - 0.097104 i  \\
$4 \cdot  10^{-5}$ &  0.293912 - 0.096706 i   &   0.295266 - 0.096741 i  \\
$5 \cdot  10^{-5}$ &  0.294166 - 0.096453 i   &   0.295776 - 0.096360 i  \\
$10^{-4}$ &         0.295543 -   0.095136 i  &    0.299002 - 0.094337 i  \\
$3 \cdot 10^{-4}$ & 0.304555 -   0.091160 i &     0.279144 - 0.059728 i \\
$5 \cdot 10^{-4}$ & 0.315643 -   0.093030 i &     0.278248 - 0.138408 i \\
$7 \cdot 10^{-4}$ & 0.323746 -   0.099017 i &     0.006171 - 0.042030 i \\
$10^{-3}$ &        0.330159 -   0.109229 i &     0.082300 - 0.106786 i \\
\hline
\end{tabular}
\caption{Quasinormal modes obtained by the time-domain integration and subsequent extraction of frequencies with the Prony method for various values of $\beta$, $\ell=1$, $M=1$. The higher order WKB data with Pad\'e approximants are evidently diverge from the correct time-domain integration results when $\beta$ grows.}
\end{table}

When the multipole number $\ell$ is large, one can find quasinormal frequencies in the analytic form. It is convenient to use $\kappa=\ell +1/2$ instead of $\ell$.
We find the location of the peak of the effective potential expanded in powers of $\kappa$ and, afterwards, to use the higher order WKB technique as prescribed in \citep{Konoplya:2023moy} in order to obtain an analytic expression for the quasinormal modes in the eikonal approximation and at the first order beyond it in the regime of small $\beta$ and $1/\ell$:
\begin{eqnarray}
r_{max} &=& 3 M -\frac{M}{3 \kappa ^2} + \beta  \left(\frac{2502656 \sqrt{2} \pi ^3}{295245 M^5 \kappa ^2}-\frac{210944 \sqrt{2} \pi ^3}{32805
   M^5}\right)  \nonumber \\
   &+& O\left(\frac{1}{\kappa^{4}}, \beta^2 \right).
\end{eqnarray}
Finally, applying the above expansion for $r_{max}$ and using higher order WKB formula we obtain
\begin{eqnarray}
\hspace*{-2em}\omega &=&\frac{\kappa }{3 \sqrt{3} M}-\frac{i \K}{3 \sqrt{3} M}+\frac{29-60 \K^2}{1296 \sqrt{3} M \kappa }+ \beta\left(\frac{13312 \sqrt{\frac{2}{3}} \pi ^3 \kappa }{59049 M^7}\right. \nonumber \\
\hspace*{-2em} &+&\left.\frac{115712 i \sqrt{\frac{2}{3}} \pi ^3
   \K}{59049 M^7} + \frac{64 \sqrt{\frac{2}{3}} \pi ^3 \left(91860\K^2-11783\right)}{7971615 M^7 \kappa
   }\right)\, \nonumber \\ 
\hspace*{-2em} &+&O\left(\frac{1}{\kappa^{2}}, \beta^2\right).
\end{eqnarray}

\begin{figure}
 \resizebox{\linewidth}{!}{\includegraphics{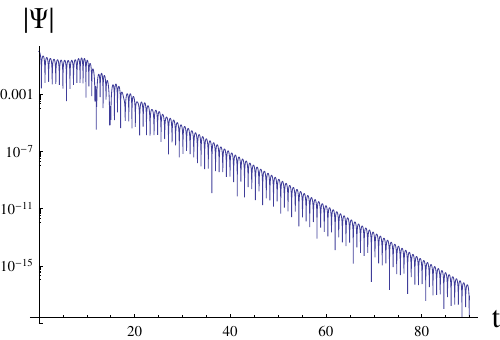}}
\caption{Time domain profile for the stable decayed oscillations $\beta = 0.002$, $M=1$, $\ell=20$. The fitting in the range formally gives $30<t<85$ gives two modes $\omega_{0} = 4.14628 - 0.431359 i$ and $\omega_{1} =6.29882 - 0.629939 i$ which is the reflection of the two stages of oscillations, the Frobenius method gives $\omega_{0} = 4.147393-0.431639 i$ and $\omega_{1}= 6.296855 - 0.627813 i$ while the WKB formula produces only the second of the two modes $\omega_{1} = 6.29683 - 0.62783 i$.}\label{figEikonal}   
\end{figure}

Neglecting orders $\sim 1/\kappa$ in the above equation, we obtain the eikonal formula which is exact in the regime $\ell \rightarrow \infty$. The eikonal formula is interesting, because, as was stated in \citep{Cardoso:2008bp},  parameters connected to unstable circular null geodesics around a spherically symmetric, and asymptotically flat or de Sitter black hole are dual to the quasinormal modes in the $\ell \gg n$ limit.  Thus,  real and imaginary parts of the $\ell \gg n$ quasinormal mode are proportional to the frequency and instability timescale of the circular null geodesics, respectively:
\begin{equation}\label{QNM}
\omega_n=\Omega\ell-\imo(n+1/2)|\lambda|, \quad \ell \gg n.
\end{equation}
Here $\Omega$ is the angular velocity at the unstable null geodesics, and $\lambda$ is the Lyapunov exponent.

An intricate aspect of this correspondence lies in its applicability, which extends to the Schwarzschild black hole and several other scenarios. However, in \citep{Konoplya:2017wot}, it was demonstrated that this correspondence falters when the wave-like equation's dominant eikonal centrifugal term deviates from $f(r) \ell (\ell +1)/r^2$. This divergence occurs, for instance, in the context of gravitational perturbations within the Einstein-Gauss-Bonnet \citep{Konoplya:2017wot,Konoplya:2020bxa} or Einstein-dilaton-Gauss-Bonnet theories \citep{Konoplya:2019hml}. Broadly speaking, the aforementioned correspondence holds true when the first-order WKB formula \citep{Schutz:1985km} is applicable, signifying the presence of an effective potential with a singular maximum and two turning points.

Furthermore, even within these scenarios, the WKB formula might prove inadequate in reproducing the complete spectrum of the eikonal regime. This limitation becomes evident in the Schwarzschild-de Sitter spacetime, where two categories of modes emerge: Schwarzschild modes that are influenced by the $\Lambda$-term \citep{Zhidenko:2003wq,Konoplya:2004uk}, and de Sitter modes that are modified by the presence of a black hole \citep{Konoplya:2022xid,Cardoso:2017soq}. Consequently, as elucidated in \citep{Konoplya:2022gjp}, although the correspondence suggested by equation (\ref{QNM}) holds formally, the parameter $n$ no longer accurately represents the overtone number.

Here we observe, in a sense, similar situation: as can be seen from fig. \ref{figEikonal}, there are two stages of the ringing on which two different modes dominate. The mode at the first stage is slower decaying and, formally, is the fundamental one, while the first overtone dominates at the second longer stage. However, the fundamental mode which is reproduced  with the time-domain integration and Frobenius method with very good concordance between them, cannot be found by the WKB method which gives only the first overtone. Thus, while the correspondence works for the part of the eikonal spectrum, it does not allow one to reproduce the whole spectrum in the eikonal regime. It is essential that the effective potential at such small values of $\beta$ appears as a WKB-good single peak barrier.  It is worth mentioning that consequently the break-down of the null geodesics/eikonal quasinormal modes correspondence for our case must also violate the correspondence between the quasinormal modes and radius of the shadow considered in \citep{Jusufi:2019ltj,Jafarzade:2020ova}.
One should also remember that null geodesics not always determine the trajectories for photon movement. Thus in the non-linear electrodynamics, photons do not move along the null geodesics \cite{Chen:2018vuw,Chen:2019dip,Toshmatov:2019gxg}.

Lower modes are shown in tables I and II, where we can see that the WKB approach cannot be trusted unless the parameter $\beta$ is tiny. At the same time, the time-domain integration always provides relatively good accuracy for the fundamental mode with relative error less than one percent at least for $\ell=1$ when there are sufficient number of oscillations in the profile. Nevertheless, it is difficult to achieve great accuracy with the time-domain integration, because of arbitrariness of the fitting of the profile with a sum of exponents within the Prony method.

\vspace{4mm}
\section{Conclusions}
This work is the first analysis of quasinormal modes of the black hole in the Starobinsky-Bel Robinson gravity suggested recently in \citep{CamposDelgado:2022sgc}. It proved out that the black hole possesses a rich quasinormal spectrum with the following features: 
\begin{enumerate}
\item At unphysically large values of the parameter of compactification $\beta$ it produces instability, while the regime of small $\beta$ is free from instability
\item In the eikonal limit $\ell \rightarrow \infty$  the correspondence breaks down, because it reproduces only part of the eikonal spectrum, but, unlike \citep{Konoplya:2022gjp}, this breakdown is for an asymptotically flat space.\\
\item The ringing consists of the two stages at each of which different modes dominate. This could be well seen in the regime of large $\ell$.
\end{enumerate}

Our work could be extended to fields of other spin and considerations of higher overtones, which could be done with the help of the Frobenius method. 

\vspace{5mm}
\section*{Acknowledgements}
The author would like to acknowledge R. A. Konoplya for careful reading of the manuscript and most useful discussions. This work was supported by RUDN University research project FSSF-2023-0003.


\bibliographystyle{elsarticle-harv} 
\bibliography{BelRobinsonStar1}
\end{document}